\documentclass[10pt,conference]{IEEEtran}
\usepackage{amsfonts}
\usepackage{amsfonts}
\usepackage{setspace}

\usepackage{amsmath}
\usepackage{amssymb}
\usepackage[dvips]{graphicx}
\usepackage{epsfig}
\usepackage{hyperref}

\usepackage{amsmath}
\usepackage{amssymb}
\usepackage[dvips]{graphicx}
\usepackage{epsfig}

\newcommand{\tsnr}{{\text{\footnotesize{SNR}}}}

\newcommand{\E}{\mathbb{E}}

\newcommand{\C}{{\sf{C}}}

\newcommand{\Pb}{\bar{P}}

\newcommand{\figsize}{0.5}

\begin{document}

%
\title{Secure Communication over Fading Channels with Statistical QoS Constraints}



%
\author{\authorblockN{Deli Qiao, Mustafa Cenk Gursoy, and Senem
Velipasalar}
\authorblockA{Department of Electrical Engineering\\
University of Nebraska-Lincoln, Lincoln, NE 68588\\ Email:
dqiao726@huskers.unl.edu, gursoy@engr.unl.edu,
velipasa@engr.unl.edu} }


\maketitle

\begin{abstract}\footnote{This work was supported by the National Science Foundation under Grants CNS--0834753, and CCF--0917265.}
In this paper, secure transmission of information over an ergodic
fading channel is studied in the presence of statistical quality of
service (QoS) constraints.
We employ effective capacity 
 to measure the secure
throughput of the system, i.e., \emph{effective secure throughput}.
We assume that the channel side information (CSI) of the main
channel is available at the transmitter side. Under different
assumptions on the availability of the CSI of the eavesdropper
channel, we investigate the optimal power control policies that
maximize the \emph{effective secure throughput}. In particular, when
the CSI of the eavesdropper channel is available at the transmitter,
it is noted that 
opportunistic transmission is no longer optimal and the transmitter
should not wait to send the data at a high rate until the main
channel is much better than the eavesdropper channel. Moreover, it
is shown that the benefits of the CSI of the eavesdropper channel
diminish as QoS constraints become more stringent.

\end{abstract}

\section{Introduction}
Security is an important issue in wireless systems due to the
broadcast nature of wireless transmissions. In a pioneering work, Wyner in \cite{wyner} addressed the security problem from an information-theoretic point of view and considered a wire-tap
channel model. He proved that secure transmission of
confidential messages to a destination in the presence of a
degraded wire-tapper can be achieved, and he established the secrecy capacity
which is defined as the highest rate of reliable communication from the
transmitter to the legitimate receiver while keeping the wire-tapper
completely ignorant of the transmitted messages. Recently, there has
been numerous studies addressing information theoretic
security. For instance, the impact of
fading has been investigated in \cite{lai}, where it has been shown
that a non-zero secrecy capacity can be achieved even when the
eavesdropper channel is better than the main channel on average. The
secrecy capacity region of the fading broadcast channel with
confidential messages and associated optimal power control policies
have been identified in \cite{liangsecure}, where it is shown that the transmitter
allocates more power as the strength of the main channel increases with respect to that of the
eavesdropper channel.

In addition to security issues, providing acceptable performance and
quality is vital to many applications. For instance, voice over IP (VoIP)
and interactive-video (e.g,. videoconferencing) systems are required to satisfy certain buffer or delay constraints. In this paper, we
consider statistical QoS constraints in the form of limitations on the buffer length, and incorporate the concept
of effective capacity \cite{dapeng}, which can be seen as the
maximum constant arrival rate that a given time-varying service
process can support while satisfying statistical QoS guarantees. The
analysis and application of effective capacity in various settings
have attracted much interest recently (see e.g.,
\cite{jia}--\cite{liu-cooperation} and references therein). We
define the \emph{effective secure throughput} as the maximum
constant arrival rate that can be supported while keeping the
eavesdropper ignorant of these messages in the presence of QoS constraints.
We assume that the CSI of the main channel is available at the
transmitter side. Then, we derive the optimal power control policies
that maximize the effective secure throughput under different
assumptions on the availability of the CSI of the eavesdropper
channel. Through this analysis, we find that due to the
introduction of QoS constraints, the transmitter cannot reserve its
power for times at which the main channel is much stronger than the eavesdropper channel. Also, we note that
the CSI of the eavesdropper provides little help when QoS
constraints become more stringent.

The rest of the paper is organized as follows. Section II briefly
describes the system model and the necessary preliminaries on
statistical QoS constraints and effective capacity. In Section III,
we present our results for both full and only main CSI scenarios.
Finally, Section IV concludes the paper.

\section{System Model and Preliminaries}

\begin{figure}
\begin{center}
\includegraphics[width=0.45\textwidth]{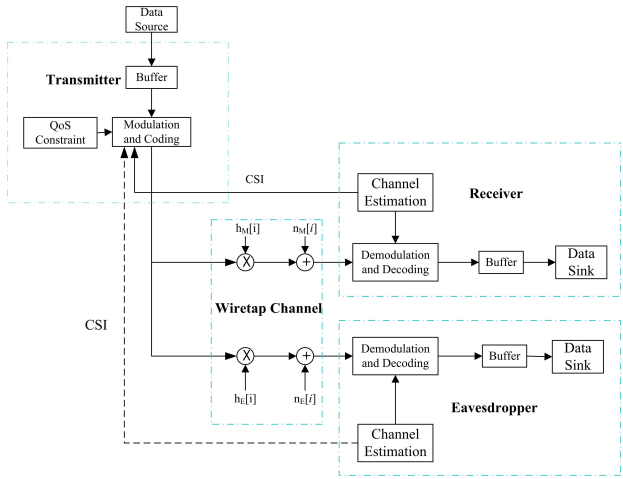}
\caption{The general system model.}\label{fig:systemmodel}
\end{center}
\end{figure}
\subsection{System Model}

The system model is shown in Fig. \ref{fig:systemmodel}. It is
assumed that the transmitter generates data sequences which are
divided into frames of duration $T$. These data frames are initially
stored in the buffer before they are transmitted over the wireless
channel. The channel input-output relationships are given by
\begin{align}
Y_1[i]&=h_1[i]X[i]+Z_1[i]\\
Y_2[i]&=h_2[i]X[i]+Z_2[i]
\end{align}
where $i$ is the frame index, $X[i]$ is the channel input in the
$i$th frame, and $Y_1[i]$ and $Y_2[i]$ represent the channel outputs
at the receivers 1 and 2 in frame $i$, respectively. We assume that
$\{h_j[i],\, j=1,2\}$'s are jointly stationary and ergodic
discrete-time processes, and we denote the magnitude-square of the
fading coefficients by $z_j[i]=|h_j[i]|^2$. Considering that
receiver 1 is the main user and receiver 2 is the
eavesdropper, we in the rest of the paper express $z_1$ and $z_2$ as $z_M$ and $z_E$, respectively, for more clarity. The channel input is subject
to an average power constraint $\E\{|X[i]|^2\} \le \Pb$, and we
assume that the bandwidth available for the system is $B$. Above, the noise component
$Z_j[i]$ is a zero-mean, circularly symmetric, complex Gaussian
random variable with variance $\E\{|Z_j[i]|^2\} = N_j$ for $j = 1,2$. The additive
Gaussian noise samples $\{Z_j[i]\}$ are assumed to form an
independent and identically distributed (i.i.d.) sequence.

We denote the average transmitted signal to noise ratio
with respect to receiver 1 as $\tsnr=\frac{\Pb}{N_1 B}$. We also denote
$P[i]$ as the instantaneous transmit power in the $i$th frame. Now, the
instantaneous transmitted SNR level for receiver 1 can be expressed as
$\mu^1[i]=\frac{P[i]}{N_1 B}$. Then, the average power constraint is equivalent to the average SNR constraint
$\E\{\mu^1[i]\}\le \tsnr$ for receiver 1. If we denote the ratio
between the noise power of the two channels as
$\gamma=\frac{N_1}{N_2}$, the instantaneous transmitted SNR level
for receiver 2 becomes $\mu^2[i]=\gamma\mu^1[i]$.

\subsection{Statistical QoS Constraints and Effective Secure Throughput}
In \cite{dapeng}, Wu and Negi defined the effective capacity as the
maximum constant arrival rate\footnote{For time-varying arrival
rates, effective capacity specifies the effective bandwidth of the
arrival process that can be supported by the channel.} that a given
service process can support in order to guarantee a statistical QoS
requirement specified by the QoS exponent $\theta$. If we define $Q$
as the stationary queue length, then $\theta$ is the decay rate of
the tail of the distribution of the queue length $Q$:
\begin{equation}
\lim_{q \to \infty} \frac{\log P(Q \ge q)}{q} = -\theta.
\end{equation}
Therefore, for large $q_{\max}$, we have the following approximation
for the buffer violation probability: $P(Q \ge q_{\max}) \approx
e^{-\theta q_{\max}}$. Hence, while larger $\theta$ corresponds to
more strict QoS constraints, smaller $\theta$ implies looser QoS
guarantees. Similarly, if $D$ denotes the steady-state delay
experienced in the buffer, then $P(D \ge d_{\max}) \approx
e^{-\theta \delta d_{\max}}$ for large $d_{\max}$, where $\delta$ is
determined by the arrival and service processes
\cite{tangzhangcross2}.

The effective capacity is given by
\begin{align}\label{eq:effectivedefi}
C(\theta)=-\frac{\Lambda(-\theta)}{\theta}=-\lim_{t\rightarrow\infty}\frac{1}{\theta
t}\log_e{\mathbb{E}\{e^{-\theta S[t]}\}} \,\quad \text{bits/s},
\end{align}
where the expectation is with respect to $S[t] =
\sum_{i=1}^{t}s[i]$, which is the time-accumulated service process.
$\{s[i], i=1,2,\ldots\}$ denotes the discrete-time stationary and
ergodic stochastic service process. We define the effective capacity
obtained when the service rate is confined by the secrecy capacity
as the \emph{effective secure throughput}.

In this paper, in order to simplify the analysis while considering
general fading distributions, we assume that the fading coefficients
stay constant over the frame duration $T$ and vary independently for
each frame and each user. In this scenario, $s[i]=T R[i]$, where
$R[i]$ is the instantaneous service rate for confidential
messages in the $i$th frame duration $[iT,(i+1)T]$. Then,
(\ref{eq:effectivedefi}) can be written as
\begin{align}
C(\theta)&=
-\frac{1}{\theta T}\log_e\mathbb{E}_{\mathbf{z}}\{e^{-\theta T
R[i]}\}\,\quad \text{bits/s}, \label{eq:effectivedefirate}
\end{align}
where $R[i]$ in general depends on the fading magnitudes
$\mathbf{z} = (z_M, z_E)$. (\ref{eq:effectivedefirate}) is obtained using the
fact that instantaneous rates $\{R[i]\}$ vary independently over different frames.
The \emph{effective secure throughput} normalized by bandwidth $B$
is
\begin{equation}\label{eq:normeffectivedefi}
\C(\theta)=\frac{C(\theta)}{B} \quad \text{bits/s/Hz}.
\end{equation}

\section{Secrecy Capacity with QoS Constraints}
Gopala \emph{et al.} in \cite{lai}
investigated the secrecy capacity in ergodic fading channels without
delay constraints. They considered two cases: full CSI at the
transmitter and only main CSI at the transmitter. In this section, we also
consider these two cases but in the presence of statistical QoS constraints.

\subsection{Full CSI at the Transmitter}\label{sec:fullcsi}

In this part, we assume that the perfect CSI of the main channel and the eavesdropper channel is available at the transmitter. The
transmitter is able to adapt the transmitted power according to the
instantaneous values of $z_M$ and $z_E$ only when $z_M>\gamma z_E$.
The secrecy capacity is then given by
\begin{align}
\hspace{-.3cm}R_s=\left\{\begin{array}{ll}
\log_2(1+\mu(z_M,z_E)z_M)
\\
-\log_2(1+\gamma\mu(z_M,z_E)z_E),\,&z_M>\gamma
z_E
\\0,\,&\text{else.}\end{array}\right.
\end{align}
where $\mu(z_M,z_E)$ is the optimal power allocated when $z_M$ and
$z_E$ are known at the transmitter.

In the presence of QoS constraints,
the optimal power allocation policy in general depends on the QoS exponent
$\theta$\footnote{Due to this dependence, we henceforth use
$\mu(\theta,z_M,z_E)$ to denote the power allocation policy under QoS
constraints.}. Hence, the secure throughput can be expressed as
\begin{small}
\begin{align}
\hspace{-.9cm}\C_E(\theta)&=\hspace{-.65cm}\max_{\substack{\mu(\theta,z_M,z_E) \\\E\{\mu(\theta,z_M,z_E)\}\le\tsnr}}-\frac{1}{\theta
TB}\log_e\Bigg(\int_0^\infty\int_0^{z_E}p_{z_M}(z_M)p_{z_E}(z_E)dz_Mdz_E\nonumber\\
&+\int_0^\infty\int_{z_E}^{\infty}\left(\frac{1+\mu(\theta,z_M,z_E)
z_M}{1+\gamma\mu(\theta,z_M,z_E)
z_E}\right)^{-\beta}p_{z_M}(z_M)p_{z_E}(z_E)dz_Mdz_E\Bigg).\nonumber\\
&\phantom{\max_{\mu(\theta,z_M,z_E)}} 
\label{eq:fullcsimaxp}
\end{align}
\end{small}
Note that the first term in the $\log$ function is a constant, and
$\log$ is a monotonically increasing function.
The maximization problem in (\ref{eq:fullcsimaxp}) is equivalent to the following minimization
problem
\begin{align}
&\min_{\substack{\mu(\theta,z_M,z_E)\\\E\{\mu(\theta,z_M,z_E)\}\le\tsnr}}\int_0^\infty\int_{z_E}^{\infty}\left(\frac{1+\mu(\theta,z_M,z_E)
z_M}{1+\gamma\mu(\theta,z_M,z_E)
z_E}\right)^{-\beta}\nonumber\\
&\hspace{3cm}\times p_{z_M}(z_M)p_{z_E}(z_E)dz_Mdz_E. 
\label{eq:fullcsimaxprev}
\end{align}
It is easy to check that when $z_M>\gamma z_E$
\begin{equation}
f(\mu)=\left(\frac{1+\mu z_M}{1+\gamma \mu z_E}\right)^{-\beta}
\end{equation}
is a convex function in $\mu$. According to \cite{convex},
non-negative integral preserves convexity, hence the objective
function is convex in $\mu$. Then, we can form the following
Lagrangian function, denoted as $\mathcal{J}$:
\begin{small}
\begin{align}
\hspace{-.2cm}\mathcal{J}&=\int_0^\infty\int_{z_E}^{\infty}\left(\frac{1+\mu(\theta,z_M,z_E)
z_M}{1+\gamma \mu(\theta,z_M,z_E)
z_E}\right)^{-\beta}p_{z_M}(z_M)p_{z_E}(z_E)dz_Mdz_E\nonumber\\
&+\lambda\left(\int_0^\infty\int_{z_E}^{\infty}\mu(\theta,z_M,z_E)p_{z_M}(z_M)p_{z_E}(z_E)dz_Mdz_E-\tsnr\right)
\end{align}
\end{small}
Taking the derivative of the Lagrangian function over
$\mu(\theta,z_M,z_E)$, we get the following optimality
condition:
\begin{align}
&\frac{\partial \mathcal{J}}{\partial
\mu(\theta,z_M,z_E)}=\lambda-\beta\left(\frac{1+\mu(\theta,z_M,z_E)
z_M}{1+\gamma \mu(\theta,z_M,z_E)
z_E}\right)^{-\beta}\nonumber\\
&\phantom{=-\beta}\times\frac{z_M-\gamma z_E}{(1+\mu(\theta,z_M,z_E)
z_M)(1+\gamma \mu(\theta,z_M,z_E) z_E)}=0\label{eq:fullcsioptcond}
\end{align}
where $\lambda$ is the Lagrange multiplier whose value is chosen to satisfy the average power
constraint with equality. For any channel state pairs $(z_M,z_E)$,
$\mu(\theta,z_M,z_E)$ can be obtained from the above condition.
Whenever the value of $\mu(\theta,z_M,z_E)$ is negative, it
follows from the convexity of the objective function with respect to
$\mu(\theta,z_M,z_E)$ that the optimal value of
$\mu(\theta,z_M,z_E)$ is 0.

There is no closed-form solution to (\ref{eq:fullcsioptcond}).
However, since the left-hand side of (\ref{eq:fullcsioptcond}) is a
monotonically increasing concave function, numerical techniques such
as bisection search method can be efficiently adopted to
derive the solution. 

The secure throughput can be determined by substituting the optimal
power control policy for $\mu(\theta,z_M,z_E)$ in
(\ref{eq:fullcsimaxp}). Exploiting the optimality condition in
(\ref{eq:fullcsioptcond}), we can notice that when
$\mu(\theta,z_M,z_E)=0$, we have $z_M-\gamma
z_E=\frac{\lambda}{\beta}$. Meanwhile,
\begin{align}
&\left(\frac{1+\mu(\theta,z_M,z_E) z_M}{1+\gamma\mu(\theta,z_M,z_E)
z_E}\right)^{-\beta} \nonumber
\\
&\times \frac{1}{(1+\mu(\theta,z_M,z_E) z_M)(1+\gamma
\mu(\theta,z_M,z_E) z_E)}<1.
\end{align}
Thus, we must have $z_M-\gamma
z_E>\frac{\lambda}{\beta}$ for $\mu(\theta,z_M,z_E)>0$, i.e.,
$\mu(\theta,z_M,z_E)=0$ if $z_M-\gamma z_E\le\frac{\lambda}{\beta}$.
Hence, we can write the secure throughput as
\begin{align}
\C_E(\theta)&=-\frac{1}{\theta
TB}\log_e\Bigg(\int_0^\infty\int_0^{\gamma z_E+\frac{\lambda}{\beta}}p_{z_M}(z_M)p_{z_E}(z_E)dz_Mdz_E\nonumber\\
&+\int_0^\infty\int_{\gamma
z_E+\frac{\lambda}{\beta}}^{\infty}\left(\frac{1+\mu(\theta,z_M,z_E)
z_M}{1+\gamma \mu(\theta,z_M,z_E) z_E}\right)^{-\beta}\nonumber\\
&\hspace{3cm}\times p_{z_M}(z_M)p_{z_E}(z_E)dz_Mdz_E\Bigg)
\end{align}
where $\mu(\theta,z_M,z_E)$ is the derived optimal power control
policy.

\subsection{Only Main Channel CSI at the Transmitter}

In this section, we assume that the transmitter has only the CSI of
the main channel (the channel between the transmitter and the
legitimate receiver). Under this assumption, it is shown in
\cite{lai} that the secrecy rate for a specific channel state pair
becomes
\begin{align}
R_s=\left[\log_2(1+\mu(z_M)z_M)-\log_2(1+\gamma \mu(z_M)z_E)\right]^+
\end{align}
where $\mu(z_M)$ is the optimal power allocated when only $z_M$ is known
at the transmitter.

In this case, the secure throughput can be expressed as
\begin{small}
\begin{align}
\C_E(\theta)&=\hspace{-.5cm}\max_{\substack{\mu(\theta,z_M)\\\E\{\mu(\theta,z_M)\}\le\tsnr}}-\frac{1}{\theta
TB}\log_e\Bigg(\int_0^\infty\int_{z_M}^{\infty}p_{z_M}(z_M)p_{z_E}(z_E)dz_Edz_M\nonumber\\
&+\int_0^\infty\int_{0}^{z_M}\left(\frac{1+\mu(\theta,z_M)
z_M}{1+\gamma \mu(\theta,z_M)
z_E}\right)^{-\beta}p_{z_M}(z_M)p_{z_E}(z_E)dz_Edz_M\Bigg)
. \label{eq:maincsimaxp}
\end{align}
\end{small}
Similar to the discussion in Section \ref{sec:fullcsi}, we get the following
equivalent minimization problem:
\begin{align}
\min_{\substack{\mu(\theta,z_M)\\\E\{\mu(\theta,z_M)\}\le\tsnr}}\int_0^\infty\int_{0}^{z_M}&\left(\frac{1+\mu(\theta,z_M)
z_M}{1+\gamma \mu(\theta,z_M)
z_E}\right)^{-\beta} \nonumber
\\
&\times p_{z_M}(z_M)p_{z_E}(z_E)dz_Mdz_E
. \label{eq:maincsimaxprev}
\end{align}
The objective function in this case is convex, and with a similar
Lagrangian optimization method, we can get the following optimality
condition:
\begin{align}
\frac{\partial \mathcal{J}}{\partial
\mu(\theta,z_M)}&=-\beta\int_{0}^{z_M}\left(\frac{1+\mu(\theta,z_M)
z_M}{1+\gamma\mu(\theta,z_M)
z_E}\right)^{-\beta-1}\nonumber\\
&\phantom{=-\beta}\times\frac{z_M-\gamma
z_E}{(1+\gamma\mu(\theta,z_M)
z_E)^2}p_{z_E}(z_E)dz_E+\lambda=0\label{eq:maincsioptcond}
\end{align}
where $\lambda$ is a constant chosen to satisfy the average power
constraint with equality. If the obtained power level
$\mu(\theta,z_M)$ is negative, then the optimal value of
$\mu(\theta,z_M)$ becomes 0 according to the convexity of the
objective function in (\ref{eq:maincsimaxprev}). Now that
non-negative integral does not change the concavity, the LHS of
(\ref{eq:maincsioptcond}) is still a monotonic increasing concave
function of $\mu(\theta,z_M)$.

\begin{figure}
\begin{center}
\includegraphics[width=\figsize\textwidth]{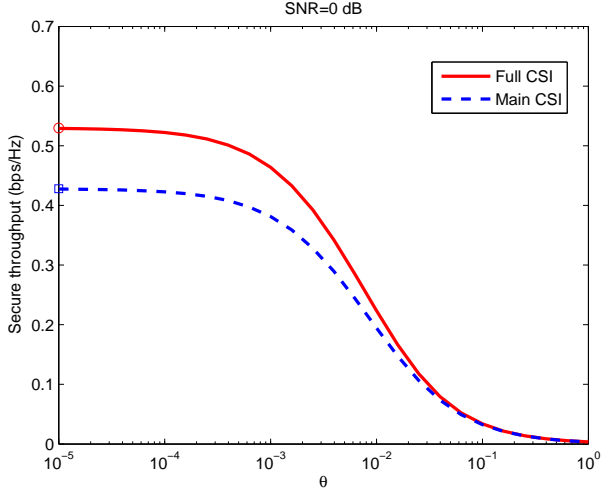}
\caption{The effective secure throughput with $\theta$ in Rayleigh
channel with $\E\{z_E\}=\E\{z_M\}=1$. $\gamma=1$.
}\label{fig:secrecym=1}
\end{center}
\end{figure}

The secure throughput can be determined by substituting the optimal
power control policy for $\mu(\theta,z_M)$ in
(\ref{eq:maincsimaxp}). Exploiting the optimality condition in
(\ref{eq:maincsioptcond}), we can notice that when
$\mu(\theta,z_M,z_E)=0$, we have
\begin{align}
-\beta\int_{0}^{z_M}(z_M-\gamma
z_E)&p_{z_E}(z_E)dz_E+\lambda=0
\\
&\Rightarrow \int_0^{z_M}P(z_E\le
t)dt=\frac{\lambda}{\beta}
\end{align}
Denote the solution to the above equation as $\alpha$. Considering that
\begin{align}
\left(\frac{1+\mu(\theta,z_M,z_E) z_M}{1+\gamma \mu(\theta,z_M,z_E)
z_E}\right)^{-\beta-1}\frac{1}{(1+\gamma \mu(\theta,z_M,z_E)
z_E)^2}<1,
\end{align}
we must have $z_M>\alpha$ for $\mu(\theta,z_M)>0$, i.e.,
$\mu(\theta,z_M)=0$ if $z_M\le\alpha$. Hence, we can write the
secure throughput as
\begin{align}
\hspace{-.4cm}\C_E(\theta)&=-\frac{1}{\theta
TB}\log_e\Bigg(\int_0^{\alpha}\int_0^\infty
p_{z_M}(z_M)p_{z_E}(z_E)dz_Edz_M\nonumber\\
&\phantom{-\frac{1}{\theta TB}}+\int_{\alpha}^\infty
\int_{z_M}^{\infty}p_{z_M}(z_M)p_{z_E}(z_E)dz_Edz_M\nonumber\\
&+\int_{\alpha}^\infty \int_0^{z_M}\left(\frac{1+\mu(\theta,z_M)
z_M}{1+\gamma \mu(\theta,z_M)
z_E}\right)^{-\beta}
\\
&\hspace{1.5cm}\times p_{z_M}(z_M)p_{z_E}(z_E)dz_Edz_M\Bigg)
\end{align}
where $\mu(\theta,z_M)$ is the derived optimal power control policy.


\subsection{Numerical Results}

\begin{figure}
\begin{center}
\includegraphics[width=\figsize\textwidth]{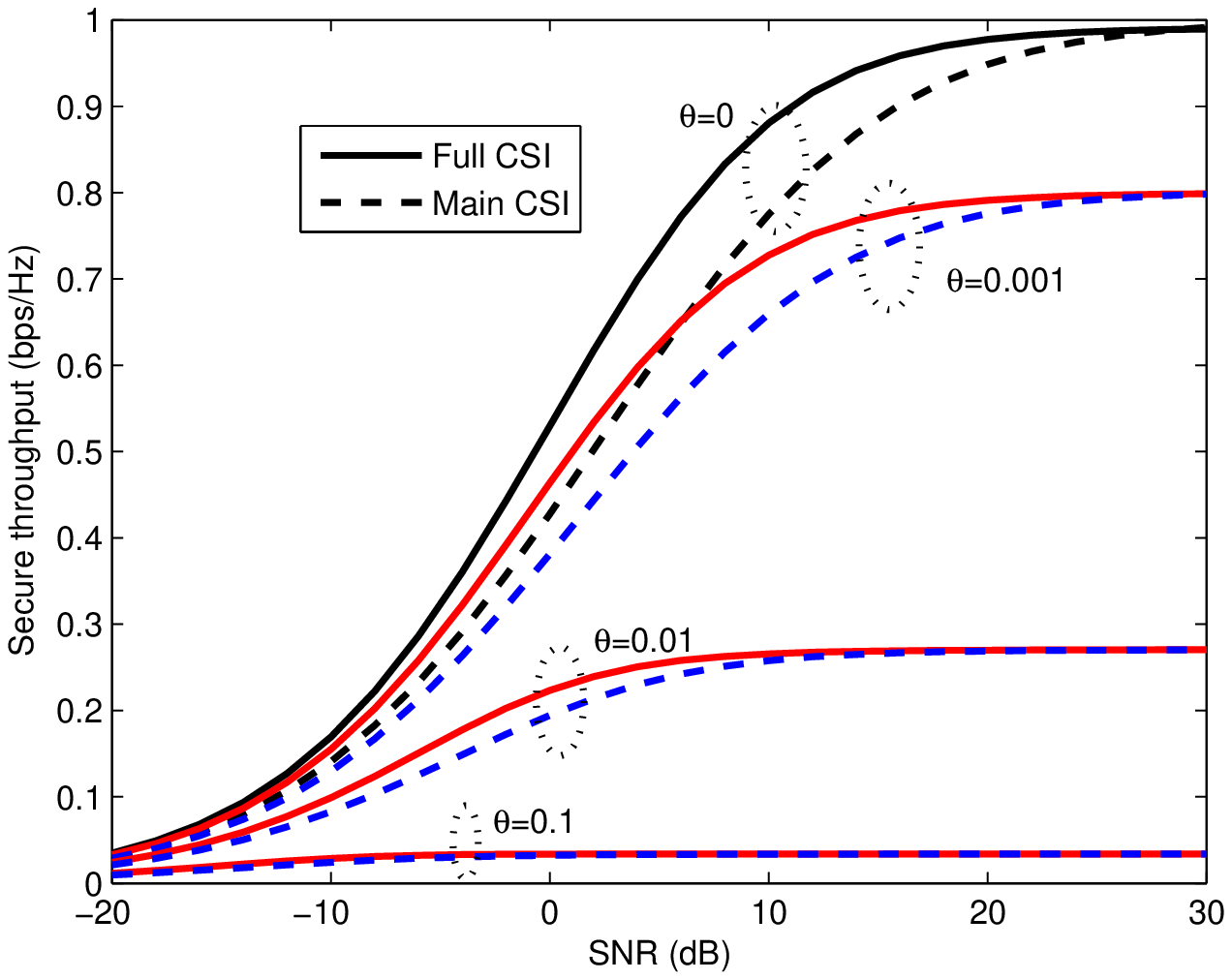}
\caption{The effective secure throughput with $\tsnr$ in Rayleigh
channel with $\E\{z_E\}=\E\{z_M\}=1$.
$\gamma=1$.}\label{fig:fixedsnr}
\end{center}
\end{figure}

In Fig. \ref{fig:secrecym=1}, we plot the effective secure
throughput as a function of the QoS exponent $\theta$ in Rayleigh
fading channel with $\gamma=1$ for the full and main CSI scenarios.
It can be seen from the figure that as the QoS constraints become
more stringent and hence as the value of $\theta$ increases, little help is provided by the CSI of the
eavesdropper channel. In Fig. \ref{fig:fixedsnr}, we plot the
effective secure throughput as $\tsnr$ varies for
$\theta=\{0,0.001,0.01,0.1\}$. Not surprisingly, we observe that the availability of the CSI of the
eavesdropper channel at the transmitter does not provide much gains in terms of increasing the
effective secure throughput in the large $\tsnr$ regime. Also, as QoS
constraints becomes more strict, we similarly note that having the CSI of the eavesdropper channel
does not increase the rate of secure transmission much even at medium $\tsnr$
levels.

\begin{figure}
\begin{center}
\includegraphics[width=\figsize\textwidth]{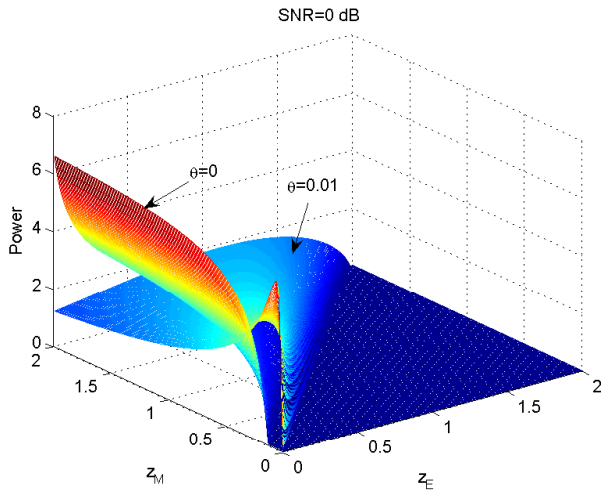}
\caption{The power allocation for full CSI scenario with $\tsnr=0$
dB in Rayleigh channel with $\E\{z_E\}=\E\{z_M\}=1$. $\gamma=1$.
}\label{fig:powerdistfull-0}
\end{center}
\end{figure}

To have an idea of the power allocation policy, we plot the power
distribution as a function of $(z_E,z_M)$ for full CSI case when $\theta=0.01$ and $\theta=0$
in Fig. \ref{fig:powerdistfull-0}. In the figure, we see that
for both values of $\theta$, no power is allocated for transmission when $z_M < z_E$ which is expected under the assumption of equal noise powers, i.e., $N_1 = N_2$. We note that when $\theta = 0$ and hence there are no buffer constraints, opportunistic transmission policy is employed. More power is allocated for cases in which the difference $z_M-z_E$ is large. Therefore, the transmitter favors the times at which the main channel is much better than the eavesdropper channel. At these times, the transmitter sends the information at a high rate with large power. When $z_M-z_E$ is small, transmission occurs at a small rate with small power. However, this strategy is clearly not optimal in the presence of buffer constraints because waiting to transmit at high rate until the main channel becomes much stronger than the eavesdropper channel can lead to buildup in the buffer and incur large delays. Hence, we do not observe this opportunistic transmission strategy when $\theta = 0.01$. In this case, we note that a more uniform power allocation is preferred. In order not to violate the limitations on the buffer length, transmission at a moderate power level is performed even when $z_M-z_E$ is small.

\section{Conclusion}
In this paper, we have anayzed the secrecy capacity in the
presence of statistical QoS constraints. We have considered the
\emph{effective secure throughput} as a measure of the performance. With
different assumptions on the availability of the full and main CSI at the transmitter,
we have investigated the associated optimal power allocation policies
that maximize the effective secure throughput. In particular, we have noted that the
transmitter allocates power more uniformly instead of concentrating its
power for the cases in which the main channel is much stronger than the eavesdropper channel. By
numerically comparing the obtained effective secure throughput, we have
shown that as QoS constraints become more stringent, the benefit of having the CSI of the
eavesdropper channel at the transmitter diminishes.

%

\end{document}